\documentclass[referee]{mn2e}

\usepackage{graphicx}
\usepackage{amsmath}

\title[X-rays and Gamma-rays from Cataclysmic Variables]
{X-rays and Gamma-rays from Cataclysmic Variables: \\
The example case of Intermediate Polar V1223 Sgr}
\author[W. Bednarek \& J. Pabich]
{W. Bednarek \& J. Pabich\\
Department of Astrophysics, University of \L \'od\'z,
ul. Pomorska 149/153, 90-236 \L \'od\'z, Poland \\
bednar@astro.phys.uni.lodz.pl; jpabich@uni.lodz.pl}
\begin{document}

\date{Accepted . Received ; in original form }

\pagerange{\pageref{firstpage}--\pageref{lastpage}} \pubyear{2007}

\maketitle

\label{firstpage}

\begin{abstract}
The accretion of matter onto intermediate polar White Dwarfs (IPWDs) seems to provide 
attractive conditions for acceleration of particles to high energies in a strongly magnetized turbulent region at the accretion disk inner radius. We consider possible
acceleration of  electrons and hadrons in such region and investigate their high energy radiation processes. It is concluded that accelerated electrons loose energy mainly on synchrotron process producing non-thermal X-ray emission. On the other hand, accelerated hadrons are convected onto the WD surface and interact with dense matter.
As a result, high energy $\gamma$-rays from decay of neutral pions and secondary leptons from decay of charged pions appear. We show that GeV-TeV $\gamma$-rays can escape from the vicinity of the WD. Secondary leptons produce synchrotron radiation in the hard X-rays and soft $\gamma$-rays. As an example, we predict the X-ray and $\gamma$-ray emission from 
IPWD V1223 Sgr. Depending on the spectral index of injected particles, this high energy emission may be detected by the ${\it Fermi}$-LAT telescope and/or the future Cherenkov Telescope Array (CTA) observatory. 
\end{abstract}

\section{Introduction}

Cataclysmic Variables (CVs) are White Dwarfs within compact binary systems which accrete matter from companion stars. The accretion process can occur in different modes depending on the strength of the surface magnetic field of the WDs, the accretion rate and the angular momentum of matter. Specific modes correspond to different types of accreting WD systems such as: polars (direct accretion onto a magnetic pole), intermediate polars (accretion onto a magnetic pole from the accretion disk), and non-magnetic WDs (the accretion disk extends to the surface of the WD), for reviews on CVs see e.g. Patterson~(1994) or Giovannelli~(2008). We are interested in possible high energy processes in the intermediate polar CVs since these objects have been recently established by the INTEGRAL as a class of the hard X-ray sources with evidences of non-thermal components (e.g. Barlow et al.~2008; Revnivtsev et al.~2008). This X-ray emission can be described either by the thermal bremsstrahlung from the accretion column close to the WD surface or by the non-thermal component whose origin is unclear.   

The physical processes in Cataclysmic Variables are expected to be similar to those observed in the X-ray binaries containing accreting neutron stars.
Therefore, CVs were also suspected to be sites of high energy processes. Unfortunately, none of such objects has been detected in the $\gamma$-rays by the EGRET telescope (Schlegel et al.~1995). However, a sporadic TeV $\gamma$-ray emission has been reported by two independent collaborations from the AE Aqr which accretes matter in the propeller phase (Meintjes et al.~1992; 1994; Bowden et al.~1992; Chadwick et al.~1995). More recent observations of this source by the Whipple and the MAGIC telescopes do not report any evidence for steady or pulsed emission (Lang et al.~1998; Sidro et al.~2008). Also a polar type CV, AM Her, has been reported  as a TeV $\gamma$-ray source by Bhat et al.~(1991). In summary, although there are not obvious evidences of $\gamma$-ray production in this type of sources their physical conditions suggest the presence of the high energy processes and encourages future observations by the satellite and ground-based observatories.

TeV $\gamma$-ray observations of AE Aqr and AM Her have stimulated investigation of 
different scenarios for the $\gamma$-ray production in CVs. They are mainly modifications of the models proposed for the high energy processes in accreting neutron stars. For the review of the older models we recommend Sect.~2 in Schlegel et al.~(1995). More recently new models have been elaborated for the possible 
TeV $\gamma$-ray emission from AE Aqr (see e.g. Kuijpers et al.~1997; Meintjes \& de Jager~2000; Ikhsanov \& Biermann~2006). In this paper we are interested in processes occurring in the intermediate polar type of CVs which operates under different accretion mode than AE Aqr.
In the case of IPWDs, the accretion process occurs through the formation of an 
accretion disk which is disrupted at the inner radius by the rotating WD magnetosphere. Starting from the disk inner radius, the accretion process occurs onto the polar region of the WD. IPWDs form specially interesting class of CVs since only they are able to emit hard X-ray radiation with non-thermal evidences. In fact, the site in which particles could be accelerated in the IPWDs might be similar in nature to that considered for the accreting neutron stars. Recently, we have discussed a model in which particles (electrons, hadrons) are accelerated in a turbulent, strongly magnetized transition region between rotating magnetosphere of the neutron star and the accretion disk (Bednarek~2009, for the earlier version see Tavani \& Liang~1993). Particles accelerated in such scenario can be responsible for the $\gamma$-ray production in the vicinity of accreting neutron stars inside Low Mass X-ray Binaries. We show that for the conditions expected in the IPWDs, particles can also reach energies suitable for the hard X-ray and GeV-TeV $\gamma$-ray production. As an example, we calculate the expected non-thermal radiation from one of the brightest IPWD, i.e. V1223 Sgr.

\section{Accretion onto White Dwarf}

White Dwarfs (WD), accreting in the accretor phase, have typical magnetic momenta two-three orders of magnitude larger then the neutron stars due to significantly larger radii. Therefore, the accretion process is influenced by the 
magnetic field typically at larger distances from the WD than in the case of neutron stars. As a result, the available gravitational energy reservoir which could be in principle converted to radiation is lower for the WDs.
The accretion rate of matter onto the WD surface, (${\dot M}_{\rm acc} = 10^{17}M_{\rm 17}$ g s$^{-1}$), can be estimated from the observed thermal X-ray emission, $L_{\rm X}$.  ${\dot M}_{\rm acc}$ and $L_{\rm X}$ can be related to the known radius and mass of the WD (we assume $R_{\rm WD}\approx 5\times 10^8$ cm and $M_{\rm WD} = 0.9M_\odot$),
\begin{eqnarray}
L_{\rm x} = GM_{\rm WD}{\dot M}/R_{\rm WD}\approx 2.4\times 10^{34}M_{\rm 17}~~~~{\rm erg~s^{-1}}.
\label{eq1}
\end{eqnarray}
\noindent 
Due to the strong magnetic field of rotating WD, the pressure of the
accreting matter is balanced by the pressure of the rotating magnetosphere
at some distance from the  WD surface. We first consider the case of quasi-spherical accretion. The distance, at which the magnetic field begins to dominate the dynamics of matter (so called the Alfven radius), can be estimated by comparing the magnetic-field energy density with the kinetic-energy density of matter (see e.g. Elsner \& Lamb~1977),
\begin{eqnarray}
B_{\rm A,o}^2/8\pi = \rho v_{\rm f}^2/2,
\label{eq2}
\end{eqnarray}
\noindent
where $B_{\rm A,o}$ is the magnetic field strength in the inner magnetosphere of the WD, $\rho = {\dot M}_{\rm acc}/(\pi R_{\rm A,O}^2v_{\rm f})$ is the density of accreting matter, $v_{\rm f} = (2GM_{\rm WD}/R_{\rm A,O})^{1/2}$ is the free fall velocity of accreting matter, $R_{\rm A,o}$ is the Alfven radius in the case of the spherically-symmetric accretion, and $G$ is the gravitational constant. The medium in the transition region is very turbulent and strongly magnetized, providing good conditions for the acceleration of particles to high energies. We estimate the location of this region from the surface of the WD by applying Eq.~(\ref{eq2}) and assuming that the magnetic field in the inner WD magnetosphere is of the dipole type, i.e. $B_{\rm A} = B_{\rm WD} (R_{\rm WD}/R_{\rm A})^3$, then
\begin{eqnarray}
R_{\rm A,o} = 3.5\times 10^{9} B_{\rm 6}^{4/7}M_{\rm 17}^{-2/7}~~~~{\rm cm},
\label{eq3}
\end{eqnarray}
\noindent
where the magnetic field at the WD surface is $B_{\rm WD} = 10^{6}B_{6}$ G. However, in the case of a disk accretion, the location of the 
Alfven radius is much more difficult to estimate since it depends on the
details of the accretion process through the disk. In general, the Alfven radius, $R_{\rm A}$, can be related to $R_{\rm A,o}$ by applying the scaling factor, $\chi$, which lays approximately in the range $\sim 0.1 - 1$ (see e.g. Lamb, Pethick \& Pines~1973),
\begin{eqnarray}
R_{\rm A} = \chi R_{\rm A,o}.
\label{eq4}
\end{eqnarray}
\noindent
Based on the known value of $R_{\rm A}$, we can estimate the magnetic field strength at the transition region,
\begin{eqnarray}
B_{\rm A} = 2.9\times 10^3\chi^{-3}M_{\rm 17}^{6/7}B_{\rm 6}^{-5/7}~~~~{\rm G}.
\label{eq5}
\end{eqnarray}
The accretion of matter onto WD can occur provided that
the rotational velocity of the magnetosphere at $R_{\rm A}$ is lower than the Keplerian velocity of the accreting matter. The rotational velocity is, 
$v_{\rm rot} = 2\pi R/P$, and the Keplerian velocity is
$v_{\rm k} = (GM_{\rm WD}/R)^{1/2}$, where $R$ is the distance from the center of the WD. The distance at which $v_{\rm rot} = v_{\rm k}$ is called the corotation radius,
\begin{eqnarray}
R_{\rm co}\approx 3.1\times 10^{9} P_2^{2/3}~~~{\rm cm},
\label{eq6}
\end{eqnarray}
\noindent
where the WD period $P = 10^2P_2$ s.
When $R_{\rm A} > R_{\rm co}$ then accretion occurs in the propeller regime.
In the opposite case, $R_{\rm A} < R_{\rm co}$, i.e. for the rotational periods of the WDs fulfilling the condition,
\begin{eqnarray}
P > 120 \chi^{3/2}B_6^{6/7}M_{17}^{-3/7}~~~{\rm s},
\label{eq7}
\end{eqnarray}
\noindent
the accretion process occurs in the accretor phase. Here we are mainly interested in the accretor phase, since in this case we can get information on the accretion rate onto the WD from the observed X-ray luminosity produced close to the WD surface.

It is expected that in the case of accretion process through the formation of an accretion disk, matter can penetrate closer to the surface of the WD. This is due to the fact that matter accumulated inside the disk can exert stronger pressure on the WD magnetosphere. Such case can be described by applying the value of the parameter $\chi < 1$. We can estimate density of matter at the disk inner radius as a function of the parameter, $\chi$, by using the balance equation,

\begin{eqnarray}
B_{\rm A}^2/8\pi = \rho v_{\rm k}^2/2.
\label{eq8}
\end{eqnarray}
\noindent
By reversing Eq.~(\ref{eq8}), we estimate density of matter at the disk inner 
radius on,

\begin{eqnarray}
\rho\approx 1.2\times 10^{13}\chi^{-5}B_6^{-6/7}M_{17}^{10/7}~~~{\rm cm^{-3}}.
\label{eq9}
\end{eqnarray}
\noindent
Note the strong dependence of density of matter in the transition region on the scaling factor $\chi$. In some cases density of matter at the disk inner radius can be a few orders of magnitude larger than in the case of spherical accretion. In the limiting case $\chi\sim 0.1$, density of matter is strong enough to provide target for the efficient interaction of relativistic hadrons already in the transition region. 
We are mainly interested in the case of $\chi\sim 1$ since the example source considered above in the paper is characterised by such penetration parameter.
However, the case with $\chi\sim 0.1$ will not require strong modifications to the discussed here scenario.

\section{Production of radiation}

We propose that accretion process onto the White Dwarf provides good conditions
for acceleration of particles and subsequent production of the high energy non-thermal radiation. Sources in which the accretion process occurs in the accretor phase are considered here in a more detail due to the best defined physical conditions. In fact, sources accreting in the propeller phase (e.g. AE Aqr) can be also characterised by similar processes as discussed below. However, in the case of propellers, it is more difficult to define the conditions at the acceleration/emission site. 

\subsection{Acceleration of particles}

In the conditions expected for the transition region (a strongly magnetized and  turbulent medium) between the inner WD magnetosphere and the accreting flow, particles should be efficiently accelerated.
The energy gain rate of particles with energy $E$ (and the Lorentz factor $\gamma$) is often parametrized by the Larmor radius of electrons and the so-called acceleration parameter,
\begin{eqnarray}
{\dot P}_{\rm acc} = \xi c E/r_{\rm L}
\approx 4.2\times 10^4\chi^{-3}\xi M_{17}^{6/7}B_{6}^{-5/7} ~~~{\rm erg~s^{-1}},
\label{eq10}
\end{eqnarray}
\noindent
where $c$ is the velocity of light, 
$r_{\rm L} = E/eB_{\rm A}$ is the Larmor radius, $B_{\rm A}$ is the magnetic field strength in the acceleration region, $e$ is the electron charge
and $\xi$ is the acceleration parameter. The acceleration parameter contains all the unknown details of the acceleration process. We estimate the order of magnitude value for $\xi$ at the transition region on, 
\begin{eqnarray}
\xi\sim \beta(v_{\rm k}/c)^2\approx 3.8\times 10^{-5}\beta\chi^{-1}M_{17}^{2/7}B_6^{-4/7}. 
\label{eq11}
\end{eqnarray}
\noindent
where $\beta\le 1$ is the factor which describes the efficiency of acceleration.

Then, the energy gains of particles can be expressed by,
\begin{eqnarray}
{\dot P}_{\rm acc} \approx 1.6\beta\chi^{-4}M_{17}^{8/7}B_{6}^{-9/7} ~~~{\rm erg~s^{-1}}.
\label{eq12}
\end{eqnarray}

In principle, both electrons and hadrons can be accelerated in such a transition region.

\subsection{Electrons}

During the acceleration process, electrons also experience energy losses due to the synchrotron radiation and the Inverse Compton Scattering (ICS) of radiation from the companion star, the accretion disk, and the surface of the WD. These energy losses determine the maximum energies of accelerated electrons, since their Larmor radius is typically far smaller than the characteristic dimensions of the considered scenario, i.e. $r_{\rm L} < R_{\rm A}$. This last condition in principle allows the acceleration of electrons up to $E_{\rm L}\approx 5\times 10^{3}\chi^{-2}M_{17}^{4/7}B_{6}^{-1/7}$ erg. It can be also easily shown, based on the formulae given below, that the escape of electrons from the acceleration region (e.g. the convection with the the accreting flow) is slower than the efficiency of the radiation processes of electrons.

The gravitational energy of accreting matter, generated in the accretion disk,
can be estimated from
\begin{eqnarray}
L_{\rm D} = {{GM_{\rm WD}{\dot M}}\over{R_{\rm A}}}\approx 3.4\times 10^{33}\chi^{-1}B_{\rm 6}^{-4/7}M_{\rm 17}^{9/7}~~~~{\rm {erg}\over{s}}.
\label{eq13}
\end{eqnarray}
In the case of a simple model of the accretion disk (Shakura \& Sunyaev~1973),
the disk luminosity can be related to its inner radius and the temperature at the inner radius by $L_{\rm D} = 4\pi R_{\rm A}^2\sigma_{\rm SB}T_{in}^4$.
Then, the temperature at the disk inner radius is,
\begin{eqnarray}
T_{\rm in} = \left({{L_{\rm D}}\over{4\pi R_{\rm A}^2\sigma_{\rm SB}}}\right)^{1/4}
\approx 2.5\times 10^{4}{{\chi^{-3/4}M_{\rm 17}^{13/28}}\over{B_{\rm 6}^{12/28}}}~~~~{\rm K}.
\label{eq14}
\end{eqnarray}
Electrons lose energy with the ICS process in the Thomson (T) and the Klein-Nishina (KN) regimes. We estimate the photon energy density at the acceleration region (the transition region at the inner disk radius) to be,
\begin{eqnarray}
\rho_{\rm r} = {{4\sigma_{\rm SB} T_{\rm in}^4}\over{c}}
\approx 3\times 10^3 \chi^{-3}M_{17}^{13/7}B_6^{-12/7}~~{\rm erg~cm^{-3}},
\label{eq15}
\end{eqnarray}
\noindent
where $\sigma_{\rm SB}$ is the Stefan-Boltzmann constant, and $T_{\rm in}$ is the inner disk temperature estimated in Eq.~(14).

We also estimate the energy density of the magnetic field at the transition region ($R_{\rm A}$ given by Eq.~(\ref{eq4})) to be,
\begin{eqnarray}
\rho_{\rm B} = B^2_{\rm A}/8\pi
\approx 3.3\times 10^5\chi^{-6}M_{17}^{12/7}B_{6}^{-10/7}~~{\rm {erg~cm^{-3}}}.
\label{eq16}
\end{eqnarray}
\noindent
The energy density of the magnetic field dominate over the energy density of radiation in the transition region, except some limitting cases when the parameter $\chi$ is close to 0.1. We do not consider such extreme model parameters here concentrating on the cases when $\chi$ is clearly above 0.1.  

The energy losses of electrons due to each of the considered process (synchrotron and IC in the T regime) can be calculated from,
\begin{eqnarray}
{\dot P}_{\rm loss} = (4/3)c\sigma_{\rm T}\rho\gamma^2\approx 2.7\times 10^{-14}\rho_{\rm B}\gamma^2~~{\rm erg~s^{-1}},
\label{eq17}
\end{eqnarray}
\noindent
where $\sigma_{\rm T}$ is the Thomson cross section. 
The maximum energies of accelerated electrons are determined by the balance between the energy gains from the acceleration process (Eq.~\ref{eq10}) and the energy losses due to synchrotron processes (Eq.~\ref{eq17}),
\begin{eqnarray}
\gamma^{\rm max}_{\rm e}\approx 1.35\times 10^4\beta^{1/2} \chi B_{6}^{1/14}M_{17}^{-2/7}.
\label{eq18}
\end{eqnarray}
Based on the above formula, we can estimate the maximum energies of synchrotron photons produced by these electrons at the transition region on,
\begin{eqnarray}
\varepsilon_{\rm max}\approx m_{\rm e}(B_{\rm A}/B_{\rm cr})(\gamma^{\rm max}_{\rm e})^2
\approx 160\xi~~~{\rm MeV},
\label{eq19}
\end{eqnarray}
\noindent
It is of the order of $\varepsilon_{\rm max}\sim 6\beta\chi^{-1}M_{17}^{2/7}B_6^{-4/7}$ keV for the value of $\xi$ estimated in Eq.~(\ref{eq11}). 

We conclude that electrons accelerated at the transition region lose energy mainly on emission of the synchrotron radiation which spectrum can extend up to hard X-rays. The ICS energy losses of electrons at the transition region can be neglected for the penetration parameter, $\chi$, not very far from unity.

\subsection{Protons}

Energy loss rate of relativistic hadrons in interactions with matter
on pion production can be estimated from,
\begin{eqnarray}
{\dot P}_{\rm pp} = \sigma_{\rm pp}c\rho Km_{\rm p}\gamma_{\rm p}\approx 8\times 10^{-6}\chi^{-5}B_6^{-6/7}M_{17}^{10/7}\gamma_{\rm p}~~~{\rm {erg}\over{s}},
\label{eq20}
\end{eqnarray}
\noindent
where the cross section for $p-p\rightarrow \pi$ collisions is $\sigma_{\rm pp} = 3\times 10^{-26}$ cm$^{2}$, $K = 0.5$ is the in-elasticity coefficient, $\rho$ is density of matter (see Eq.~\ref{eq9}), and $\gamma_{\rm p}$ is the proton Lorentz factor. The energy loss time scale can be expressed by,
\begin{eqnarray}
\tau_{\rm pp} = m_{\rm p}\gamma_{\rm p}/{\dot P}_{\rm pp}\approx
200\chi^{5}B_6^{6/7}M_{17}^{-10/7}\gamma_{\rm p}~~~{\rm s}.
\label{eq21}
\end{eqnarray}
\noindent
This interaction time for protons at the transition region can be significantly longer than the acceleration time specially for $\chi\sim 1$. 
Therefore, before efficient interactions in the acceleration region, protons may be convected onto the WD surface with the accreting matter. The characteristic free fall time scale from the distance of the Alfven radius is, 
\begin{eqnarray}
\tau_{\rm f} = R_{\rm A}/v_{\rm f}
\approx 13\chi^{3/2}B_6^{6/7}M_{17}^{-3/7}~~~{\rm s}.
\label{eq22}
\end{eqnarray}
\noindent
In such a case, the acceleration process of protons is limited by their convection from the acceleration site. We assume that the maximum energies of accelerated protons are determined by their convection out of the transition region, i.e. $\chi$ is not far from 1. In the opposite case, hadronic interactions mainly occur at the transition region and secondary leptons from decay of charged pions suffer significantly weaker magnetic field than on the surface of the WD.

By comparing the energy gain time scale, $\tau_{\rm acc} = m_{\rm p}\gamma_{\rm p}/{\dot P}_{\rm acc}$ (see Eq.~\ref{eq12}), with the escape time scale, $\tau_{\rm f}$ (Eq.~\ref{eq22}), we estimate the maximum Lorentz factors of accelerated protons,
\begin{eqnarray}
\gamma_{\rm p}\approx 1.4\times 10^4\beta\chi^{-5/2}B_6^{-3/7}M_{17}^{5/7}.
\label{eq23}
\end{eqnarray}
Protons with such Lorentz factors produce neutral pions which decay to 
two $\gamma$-ray photons. The characteristic energies of these photons can be estimated on
\begin{eqnarray}
E_\gamma\approx (K/2\mu)m_{\rm p}\gamma_{\rm p}\approx 165\beta\chi^{-5/2}B_6^{-3/7}M_{17}^{5/7}~~~{\rm GeV}.
\label{eq24}
\end{eqnarray}
\noindent
where $\mu$ ($\mu\approx 2.57\log(2\gamma_p)-6.45$ for $\gamma_{\rm p}>>1$, see Orth \& Buffington~1976) is the multiplicity of pions produced by protons with the Lorentz factor estimated in Eq.~(\ref{eq23}). So then, $\gamma$-rays from hadronic interactions could be detected by the {\it Fermi}-LAT and Cherenkov telescopes.

On the other hand, charged pions produced by protons decay into leptons which
Lorentz factors are of the order of,
\begin{eqnarray}
\gamma_{\rm e}^{\rm sec}\approx (K/4\mu)(m_{\rm p}/m_{\rm e})\gamma_{\rm p}.
\label{eq25}
\end{eqnarray}
\noindent
Leptons with such energies produce synchrotron photons in the magnetic field on the WD surface with characteristic energies,
\begin{eqnarray}
\varepsilon_{\rm max}^{\rm sec}\approx m_{\rm e}(B_{\rm WD}/B_{\rm cr})(\gamma^{\rm sec}_{\rm e})^2 \approx 38\beta^2\chi^{-5}B_6^{1/7}M_{17}^{10/7}~{\rm MeV}.
\label{eq26}
\end{eqnarray}
\noindent
The synchrotron spectra produced by secondary leptons from decay of charged pions 
are calculated assuming the value for the surface magnetic field on the WD. For likely parameters, these synchrotron spectra can extend even up to the 
${\it Fermi}$-LAT energy range.
However, in the case of efficient hadronic collisions already at the transition
region, where the magnetic field is about 2-3 orders of magnitude lower, the energies of synchrotron photons, proportional to the magnetic field strength, are also 2-3 orders of magnitude lower, i.e. these photons will be produced with energies in the hard X-rays.

\subsection{Escape of gamma-rays}

In our scenario $\gamma$-rays are produced relatively close to the surface of
the WD in the accretion column where the density of matter is large and radiation field from the whole WD surface and the accretion column is strong. We consider the escape conditions of $\gamma$-ray photons from such a radiation field.
The UV observations of the WDs allows to estimate the characteristic temperature
on their surface in the range $T_{\rm WD} = (15-25)\times 10^3$ K. Also additional component from the polar cap is sometimes observed with typical temperature up to $\sim 3\times 10^4$ K. It is  emitted from the hot spot with the area about $2-10\%$ of the WD surface (see e.g. for review de Martino~1999). We estimate 
the optical depths for $\gamma$-rays in the radiation field from the whole surface of the WD, 
\begin{eqnarray}
\tau_{\gamma\gamma}^{\rm WD}\approx R_{\rm WD}/\lambda_{\gamma\gamma}\approx R_{\rm WD}n_{\rm WD}\sigma_{\rm \gamma\gamma\rightarrow e^\pm}\approx 0.02T_4,
\label{eq27}
\end{eqnarray}
\noindent
where $n_{\rm WD}$ is density of thermal photons close to the surface of the WD with the characteristic temperature $T_{\rm WD} = 2\times 10^4T_4$ K, $\lambda_{\gamma\gamma}$ is the mean free path for $\gamma\gamma\rightarrow e^\pm$
absorption process, and $\sigma_{\rm \gamma\gamma\rightarrow e^\pm}$ is the maximum cross section for this process. 
The optical depth for $\gamma$-rays in the radiation field of the polar cap is
of similar order in the case of the  cap with the radius $R_{\rm cap}\approx 0.3R_{\rm WD}$ (upper limit), and surface temperature $T_{\rm cap} = 3\times 10^4$ (see de Martino~1999). 
Based on the above estimates, we conclude that $\gamma$-rays produced close to the surface of the WD escape without significant absorption through the thermal radiation field.

Note that the cross section for the pion production (and also $\gamma$-ray production) in hadronic collisions
is much larger than the cross sections for $\gamma$-ray absorption in the matter on creation of $e^\pm$ pairs ($\gamma+p\rightarrow e^\pm$) and pions  ($\gamma+p\rightarrow \pi$). Therefore, accelerated hadrons produce $\gamma$-rays efficiently already in the region with density which allows their free escape from the accretion column above the polar cap of the WD.

\subsection{Energetics}

The maximum power available for the acceleration of particles is limited by the energy extracted at the transition region. This energy can be supplied by two mechanisms. In the case of a quasi-spherical accretion from the stellar wind, the matter has to be accelerated to the velocity of the rotating magnetosphere at $R_{\rm A}$ or to the keplerian velocity. The rotating WD decelerates, providing energy to the turbulent region. In the case of accretion through the Lagrangian point, the matter has a large angular momentum, which has to be partially lost in the transition region in order to guarantee the accretion process up to the WD surface. The angular momentum of the accreting matter is then supplied partially to the transition region and to the WD. As a result, the WD can reach the angular momentum and accelerates. 

In the case of the accretion process occurring through the accretion disk, matter arrives to the transition region with the Keplerian velocity. 
This region is now closer to the WD surface than estimated above in $R_{\rm A}$ (see Eq.~{\ref{eq3}) by a factor $\chi\sim 0.1-1$. In order to accrete onto the WD surface, the matter from the disk has to be 
slowed down to the rotational velocity of the WD magnetosphere, i.e. from $v_{\rm kep}$ to $v_{\rm rot}$. Then, the maximum available power extracted in the transition region is,
\begin{eqnarray}
L = {{1}\over{2}}{\dot M}_{\rm acc}|v_{\rm k}^2 - v_{\rm rot}^2|,
\label{eq28}
\end{eqnarray}
\noindent
which for the accretor phase is, 
\begin{eqnarray}
L\approx 1.7\times 10^{33}\chi^{-1}B_6^{-4/7}M_{17}^{9/7}~~{\rm erg~s^{-1}}.
\label{eq29}
\end{eqnarray}
\noindent
provided that $v_{\rm k} >> v_{\rm rot}$ at the Alfven radius.
We assume that a part, $\delta$, of this energy reservoir is transferred to relativistic electrons and protons, $\delta_{\rm e}$ and $\delta_{\rm p}$, respectively.
The ratio of $\delta_{\rm e}/\delta_{\rm p}$ can not be at present reliably determined by any theory of particle acceleration. It can be only constrained by the observations interpreted in terms of a specific model.

\section{X-rays and $\gamma$-rays from V1223 Sgr}

V1223 Sgr belongs to the class of so called Intermediate Polar White Dwarfs. The conditions in  these sources allow the matter from the accretion disk to reach the surface of the star against the pressure of the rotating WD magnetosphere.
Such situation happens in the case of WDs with the intermediate surface magnetic fields, rotating with the intermediate periods (in relation to the whole population of White Dwarf class). V1223 Sgr accretes at a relatively large rate,
estimated on ${\dot M}\approx 10^{17}$ g s$^{-1}$ which is derived from the observed X-ray luminosity $L_{\rm x} = 2\times 10^{34}$ erg s$^{-1}$ (e.g. 
Barlow et al.~2006, Revnivtsev et al.~2008), assumed WD radius $4.17\times 10^8$ cm and its mass $1.17M_\odot$ (see Beuermann et al.~2004).
The distance to this source is estimated on 510 pc. The White Dwarf with the rotational period of 745.5 s is on the orbit with the period 3.366 hr around the low mass companion. The surface magnetic field of the WD is estimated
in the range $(0.5-8)\times 10^6$ G, depending on the model (see Beuermann et al.~2004).

\begin{figure}
\vskip 14.5truecm
\includegraphics{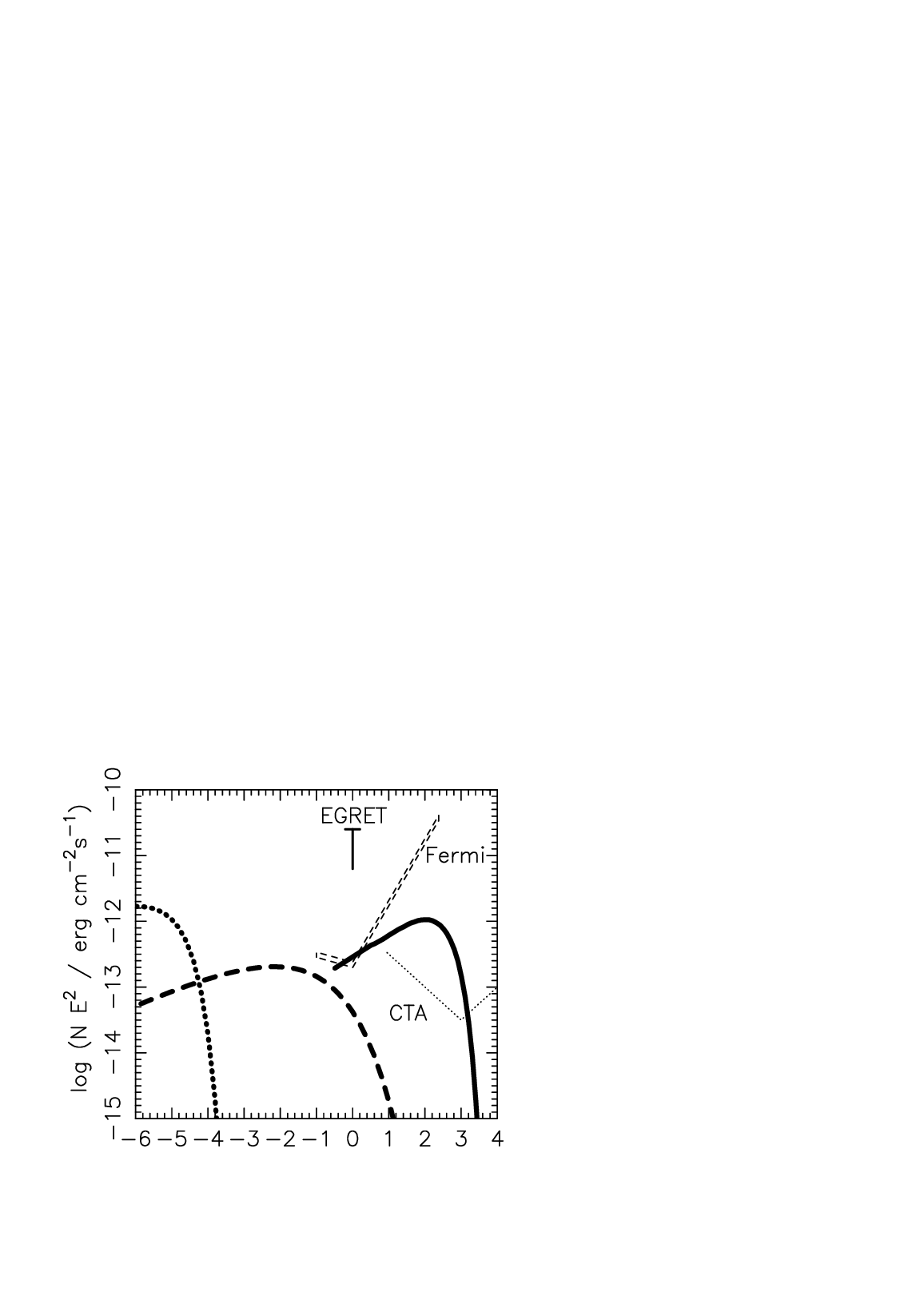}
\includegraphics{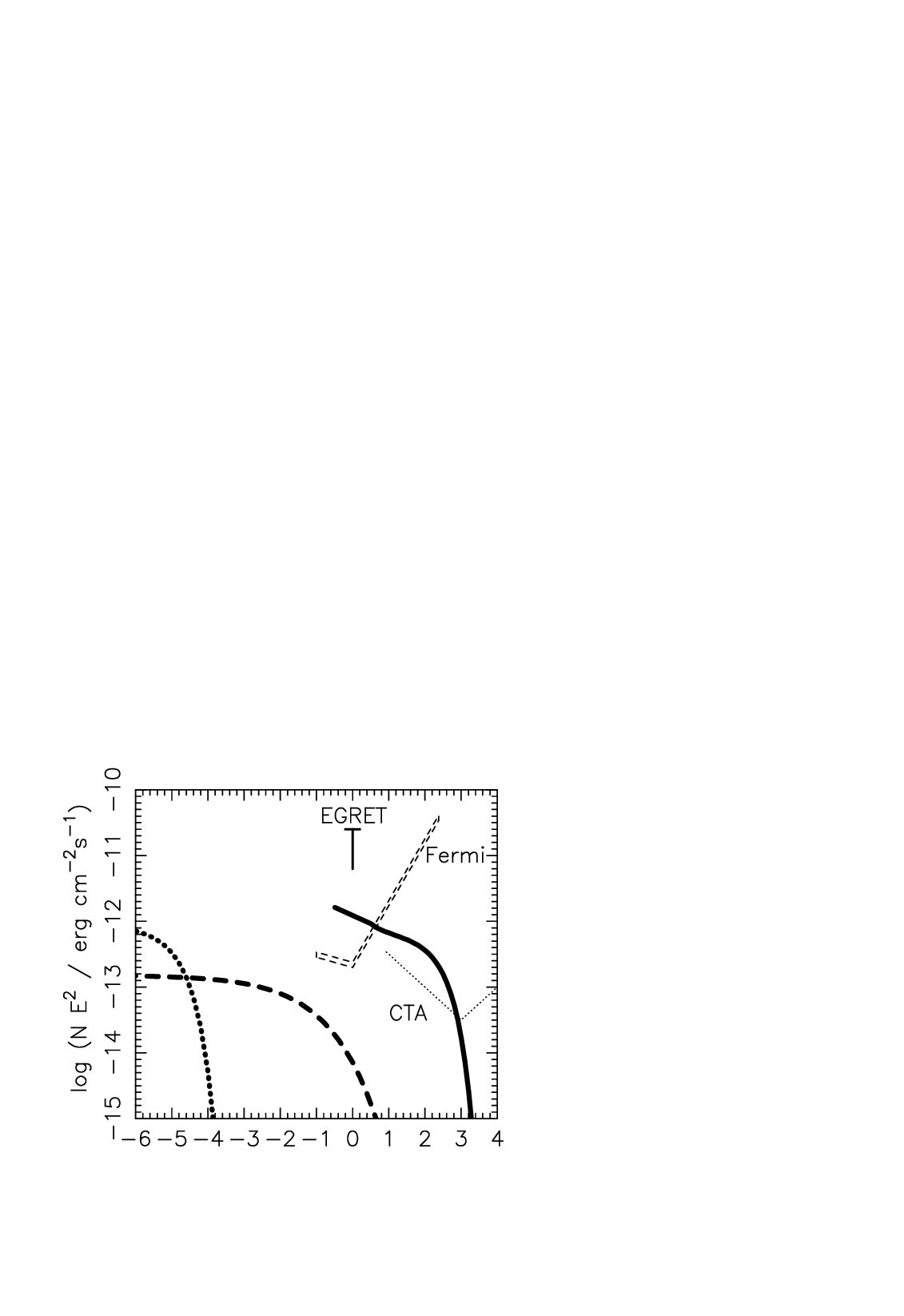}
\includegraphics{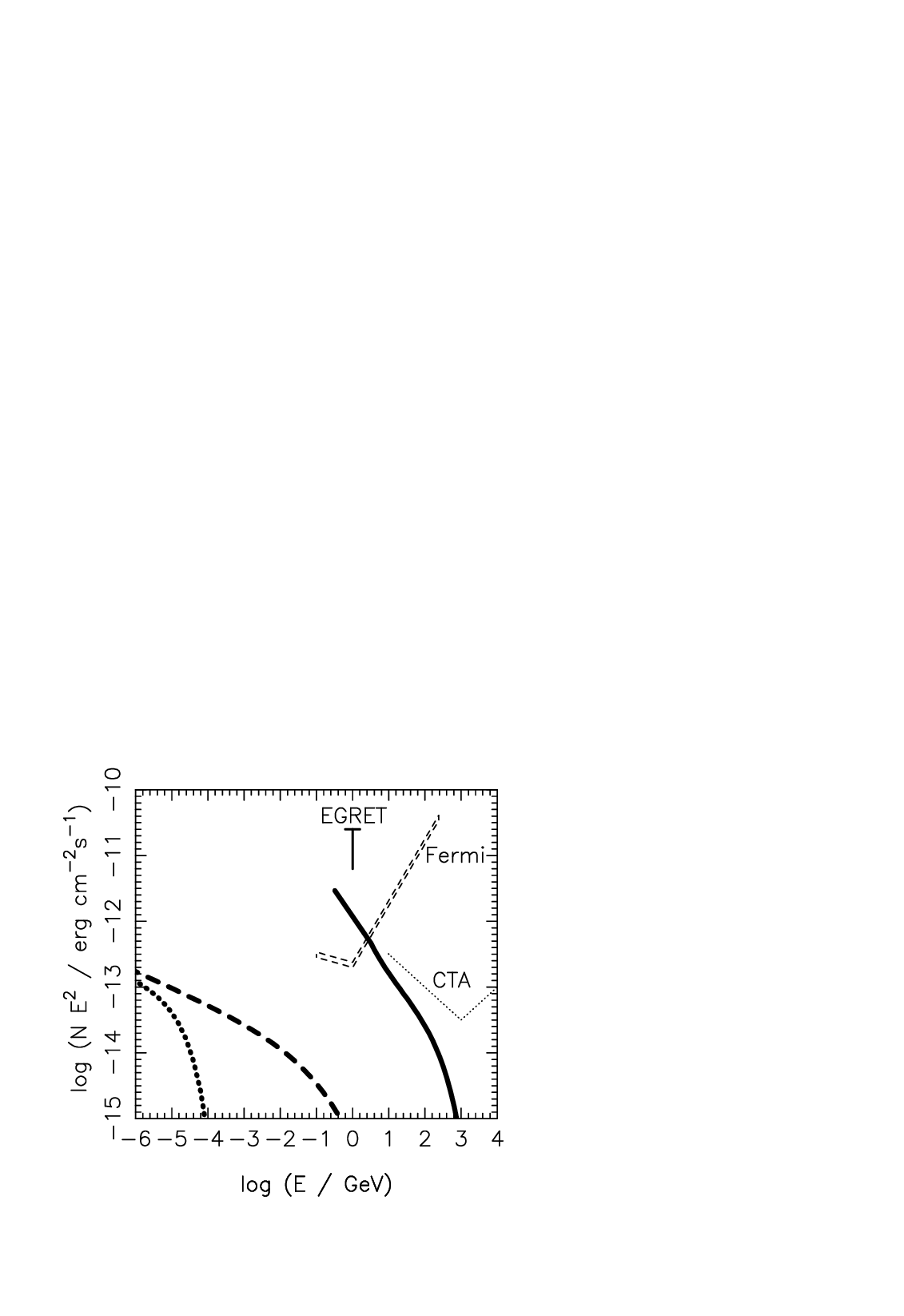}
\caption{X-ray and $\gamma$-ray spectra expected from the intermediate polar Cataclysmic Variable
V1223 Sgr. The spectra are produced by primary electrons in the synchrotron process (dotted curve), secondary leptons from decay of charged pions also in the synchrotron process (dashed) and directly from decay of neutral pions
(solid). Pions are produced by protons in hadronic collisions with matter close to the WD surface.  
Specific figures show the photon spectra for differential spectrum of the primary particles with the spectral index -1.5 (upper), -2. (middle), and -2.5 (bottom).
The energy conversion coefficients from the acceleration region into the relativistic primary electrons and protons are equal to $\delta_{\rm e} = \delta_{\rm p} = 0.1$. 
The parameters describing the WD in V1223 Sgr and the acceleration scenario are reported in the main text. The thin dashed and dotted lines mark the sensitivities of the ${\it Fermi}$-LAT telescope and the future CTA Observatory, respectively.
The EGRET upper limit on the $\gamma$-ray emission from V1223 Sgr is marked by the solid arrow.}
\label{fig1}
\end{figure}

We estimate penetration parameter into the WD magnetosphere, $\chi$, by applying the parameters of the accretion scenario for V1223 Sgr derived by Beuermann et al.~(2004, see their model A).  
Beuermann et al. consider two different models but only model A supplies the parameters for which $\gamma$-ray emission from this cataclysmic binary is possible.
Model B assumes truncation of the accretion disk at larger distances from the WD surface due to the applied much stronger surface magnetic field of the WD (model A: $B_{\rm WD} = 5\times 10^5$ G versus model B: $10^7$ G, see Beuremann et al.~2004). For such strong surface field of the WD the available energy for acceleration of particles in the transition region is significantly lower (see Eq.~\ref{eq29}). 
Therefore, detection of $\gamma$-ray emission from V1223 Sgr can decide between proposed models. For the model A, we estimate value of $\chi\approx 0.4$ based on the comparison of the derived inner radius of the accretion disk with the estimate given by our Eq.~(\ref{eq4}). The other parameters of the accretion scenario are the following: $M_{17} = 3$, $B_6 = 0.5$, in order to be consistent with the low energy observations of this binary system (see Beuermann et al.~2004).

We assume that electrons and protons are accelerated in the turbulent transition region with a power-law spectrum characterised by the exponential cut-off. The maximum energies of electrons are estimated from Eq.~(\ref{eq18}). Eventually, the spectrum of electrons can be strongly peaked at the highest possible energies due to the synchrotron energy losses during acceleration process (the so called pile-up mechanism: see e.g. 
Protheroe~2004). As shown above, electrons lose energy on different radiation processes. The most important are
the synchrotron process (dominates at the highest energies) and the ICS of thermal radiation from the WD surface. We neglect the production of $\gamma$-rays by electrons in the acceleration region in the
scattering of WD or the accretion disk radiation since its energy density can be safely neglected with respect to the energy density of the magnetic field.  
The maximum energy of protons is estimated from Eq.~(\ref{eq23}). These protons are likely convected onto the WD surface with the accreting matter and produce 
pions in collisions with matter. The spectra of $\gamma$-rays (from decay of $\pi^{\rm o}$) and secondary leptons (from decay of $\pi^\pm$) are calculated assuming the pion multiplicity production in hadronic interactions defined below Eq.~({\ref{eq24}) and the complite colling of protons on the WD surface. 

The results of calculations of the X-ray spectra from synchrotron process of primary electrons, X-rays from synchrotron process of secondary leptons and $\gamma$-rays from decay of neutral pions for different spectral indices of accelerated particles are shown in Fig.~\ref{fig1}. Note that calculated synchrotron spectra from primary electrons can not be directly observed  from V1223 Sgr in the hard X-rays since they are clearly below the emission produced in the accretion column in the thermal bremsstrahlung process. However, 
the synchrotron emission from secondary leptons extends through the soft $\gamma$-ray energy range and is close to the sensitivity limit of the ${\it Fermi}$-LAT detector. The GeV-TeV $\gamma$-ray emission from decay of neutral pions 
has been calculated for energies above 300 MeV where the sensitivity of the ${\it Fermi}$-LAT telescope is optimal. This emission could be visible by the ${\it Fermi}$-LAT and/or Cherenkov telescopes depending on the spectral index of primary particles and the efficiency of particle acceleration. There is a chance that the second stages of the H.E.S.S. and the MAGIC arrays (the H.E.S.S. II and MAGIC II) will detect some CVs of the IP type in the case of extensive observations at the sub-TeV energies.

\section{Discussion and Conclusion} 

We have calculated the non-thermal emission produced by electrons and protons which are accelerated in the transition region between the rotating WD magnetosphere and the accreting matter. The non-thermal X-ray emission from primary electrons within the transition region is clearly below the hard X-ray emission observed from the considered example Intermediate Polar Cataclysmic Variable V1223 Sgr. This observed hard X-ray emission has to have different nature, e.g. it is due to the thermal bremsstrahlung from the magnetic polar cap region of the WD surface as commonly believed. However, predicted synchrotron emission from the secondary leptons (the decay products of charged pions from hadronic collisions close to the surface of WD), extending up to GeV $\gamma$-rays, should dominate above the thermal bremsstrahlung emission at energies above $\sim$100 keV. This spectral component is a promissing target for the future hard X-ray, soft $\gamma$-ray telescopes. 

Considered here model predicts also $\gamma$-ray emission from decay of neutral pions produced in hadronic collisions in the case of V1223 Sgr which is one of the strongest X-ray emitters among the Intermediate Polar type of Cataclysmic Variables (see for details Fig.~1). Depending on the spectral index of hadrons accelerated in the transition region, the acceleration efficiency factor ($\beta$) and the ratio of the power which goes into acceleration of electrons and hadrons ($\delta_{\rm e}/\delta_{\rm p}$), this $\gamma$-ray emission can be detected by the ${\it Fermi}$-LAT telescope and/or the future CTA observatory.
At present, only ${\it Fermi}$-LAT telescope has enough sensitivity to detect
IPCVs in the most optimistic case and so constrain the values of the free parameters of the model.
In the optimal conditions, V1223 Sgr (and similar sources) have a chance to be even  detected by the extensive low threshold ($\sim$100 GeV) observations with the next generation of the Cherenkov telescope arrays such as H.E.S.S. II and MAGIC II. 
In the case of steep spectra of accelerated hadrons, the LAT telescope should be able to detect a signal from V1223 Sgr, provided that the efficiency of energy conversion from the transition region to the relativistic hadrons is equal to (or above) the applied value of $10\%$. Up to now, none of the Cataclysmic Variables has been identified in the error boxes of the ${\it Fermi}$-LAT source catalogue (Abdo et al.~2010a). We suggest the analysers to make a closer search for the CVs in the error boxes of the unidentified ${\it Fermi}$-LAT sources. 
Note that there are several similar to V1223 Sgr Intermediate Polar WDs in both hemispheres (see Tables in Barlow et al.~2006; Revnivtsev et al.~2008) which might be potentially detected in the $\gamma$-rays.

A large population of CVs have been discovered within specific globular clusters (GCs) (see e.g. Grindley et al.~2001, Pooley et al.~2002). Between them, several have been recently detected by the ${\it Fermi}$-LAT telescope in the GeV $\gamma$-rays (Abdo et al.~2010b).
Although it is rather accepted that $\gamma$-ray emission from GCs is due to the population of millisecond pulsars, it is not excluded that part of this
emission can also come from other sources, e.g. such as IPWDs considered in this paper. This possibility should be investigated in the future.
 
We discussed the model for the acceleration of particles and production of high energy radiation in the magnetosphere of accreting White Dwarf considering only the accretor phase. We expect that hadronic processes which turn to the production of GeV-TeV $\gamma$-rays might also occur in the case of White Dwarfs accreting in the propeller phase (which is probably the case of the well known Cataclysmic Variable AE Aqr, see e.g. Meintjes \& de Jager~2000). However, this accretion phase can not be constrained by the X-ray observations. Even the accretion rate can not be estimated reliably in this phase since the matter does not reach the surface of the WD. Therefore, the acceleration process of particles and their radiation processes are difficult to follow quantitatively in sources of the AE Aqr type. 
 
Note that $\gamma$-ray production is accompanied by neutrinos in the hadronic processes discussed in this paper. However, predicted fluxes of neutrinos  (which are on comparable level to the fluxes of $\gamma$-rays from decay of neutral pions) are clearly lower than the sensitivity limit of the IceCube detector.

\section*{Acknowledgments}
This work is supported by the Polish MNiSzW grant N N203 390834 and NCBiR grant
ERA-NET-ASPERA/01/10.


\label{lastpage}

\begin{thebibliography}{99}

\bibitem{ab10} Abdo, A.A. et al. 2010a ApJS 188, 405
\bibitem{ab10} Abdo, A.A. et al. 2010b A\&A, submitted
\bibitem{bar06} Barlow, E.J. et al. 2006 MNRAS 372, 224
\bibitem{bed09} Bednarek, W. 2009 A\&A 495, 919 
\bibitem{beur04} Beuermann, K., Harrison, Th. E., McArthur, B.E., Benedict, G.F., G\"ansicke, B.T. 2004 A\&A 419, 291
\bibitem{bha91} Bhat, C.L. et al. 1991 ApJ 369, 475
\bibitem{bo92} Bowden, C.C.G. et al. 1992 Aph 1, 47
\bibitem{cha95} Chadwick, P.M. et al. 1995 Aph 4, 99
\bibitem{de99} de Martino, D. 1999, MmSAI 70, 547
\bibitem{el77} Elsner, R.F., Lamb, F.K. 1977 ApJ 215, 897
\bibitem{lamb73} Lamb, F.K., Pethick, C.J., Pines, D. 1973 ApJ 184, 271
\bibitem{el77} Elsner, R.F., Lamb, F.K. 1977, ApJ 215, 897
\bibitem{gio08} Giovannelli, F. 2009 Chin. J. Astron. Astrophys. 8, 237
\bibitem{gri01} Grindley, J.E., Heinke, C., Edmonds, P.D., Murray, S.S. 2001 Science 292, 2290
\bibitem{ikh06} Ikhsanov, N.R., Biermann, P.L. 2006 A\&A 445, 305
\bibitem{kui97} Kuijpers, J. et al. 1997 A\&A 322, 242
\bibitem{la98} Lang, M.J. et al. 1998 Aph 9, 203
\bibitem{mei94} Meintjes, P.J., de Jager, O.C. 2000 MNRAS 311, 611
\bibitem{mei92} Meintjes, P.J. et al. 1992 ApJ 401, 325
\bibitem{mei94} Meintjes, P.J. et al. 1994 ApJ 434, 292
\bibitem{ob76} Orth, C.D., Buffington, A. 1976 ApJ 206, 312
\bibitem{pa94} Patterson, J. 1994, PASP 106, 209
\bibitem{poo02} Pooley, D. et al. 2002 ApJ 569, 405 
\bibitem{rev08} Revnivtsev, M., Lutovinov, A., Suleimanov, V., Sunyaev, R., Zheleznyakov, V. 2004 A\&A 426, 253
\bibitem{rev08} Revnivtsev, M., Sazonov, S., Krivonos, R., Ritter, H., Sunyaev, R. 2008 A\&A 489, 1121
\bibitem{pro04} Protheore, R.J. 2004 Aph 21, 415
\bibitem{sch95} Schlegel, E.M., Barrett, P.E., de Jager, O.C., Chanmugam, G., Hunter, S., Mattox, J. 1995 ApJ 439, 322	
\bibitem{ss73} Shakura,N.I., Sunyaev, R.A. 1973 A\&A 24, 337
\bibitem{si08} Sidro, N. et al. 2008 Proc. 30th ICRC (Merida), eds. Caballero R. et al. v. 2, p. 715
\bibitem{tav93} Tavani, M. Liang, E. in Proc. Compton Gamma-Ray Observatory Symp. ed. M. Friedlander, N. Gehrels, D. Macomb (New York: AIP), No. 280, p. 428
\end{thebibliography}
\end{document}